# Breakthrough capability for the NASA Astrophysics Explorer Program: Reaching the darkest sky

Matthew A. Greenhouse*[a], Scott W. Benson[b], Robert D. Falck[b], Dale J. Fixsen[c], Jonathan P. Gardner[a], James B. Garvin[a], Jeffery W. Kruk[a], Stephen R. Oleson[b], Harley A. Thronson[a]

[a]NASA Goddard Space Flight Center, Greenbelt, MD USA 20771; [b]NASA Glenn Research Center, Cleveland, OH USA 44135; c University of Maryland, College Park, Maryland, USA 20742

**ABSTRACT**

We describe a mission architecture designed to substantially increase the science capability of the NASA Science Mission Directorate (SMD) Astrophysics Explorer Program for all AO proposers working within the near-UV to far-infrared spectrum. We have demonstrated that augmentation of Falcon 9 Explorer launch services with a 13 kW Solar Electric Propulsion (SEP) stage can deliver a 700 kg science observatory payload to extra-Zodiacal orbit. This new capability enables up to ~13X increased photometric sensitivity and ~160X increased observing speed relative to a Sun-Earth L2, Earth-trailing, or Earth orbit with no increase in telescope aperture. All enabling SEP stage technologies for this launch service augmentation have reached sufficient readiness (TRL-6) for Explorer Program application in conjunction with the Falcon 9. We demonstrate that enabling Astrophysics Explorers to reach extra-zodiacal orbit will allow this small payload program to rival the science performance of much larger long development time systems; thus, providing a means to realize major science objectives while increasing the SMD Astrophysics portfolio diversity and resiliency to external budget pressure. The SEP technology employed in this study has strong applicability to SMD Planetary Science community-proposed missions. SEP is a stated flight demonstration priority for NASA's Office of the Chief Technologist (OCT). This new mission architecture for astrophysics Explorers enables an attractive realization of joint goals for OCT and SMD with wide applicability across SMD science disciplines.

## 1. INTRODUCTION

When ground-based astronomers want to observe the faintest objects, they do so during the new moon when the sky is at its darkest. They call this period "dark time", and telescope schedulers reserve it for the most challenging and important projects. In contrast, it has never been "dark time" for space astronomers. However, that could soon change as a result of new space propulsion capability that NASA's Science Mission Directorate (SMD) and Office of Chief Technologist (OCT) are developing to a point that it is now ready to enable a new era of low cost science missions with unprecedented capability.

The Earth is embedded in a giant cloud of dust[1]. This cloud occupies a disk-shaped region of interplanetary space encompassing much of the inner solar system. It extends from approximately the orbit of Venus to the asteroid belt, and its thickness extends roughly 0.5 AU above and below the ecliptic plane. This interplanetary dust cloud (IPD) produces a background light that, like the unwanted light of a full moon, limits the sensitivity of all space observatories that have operated within the near-UV to far-infrared spectrum. However, this limitation may soon be overcome, as NASA's Evolutionary Xenon Thruster (NEXT) and other SEP technologies are used to enable small astrophysics observatories to travel outside the IPD and reach the darkest sky for the first time.

In this paper, we describe a mission architecture study that was conducted to determine if and how a technically ready solar electric propulsion (SEP) system, such as NEXT (Figure 1), can be used to enable astrophysics Explorer missions to travel outside the IPD in order to realize an enormous gain in science capability. This study, known as the Extra Zodiacal Explorer (EZE), has shown that, by using electric propulsion to access a dark sky orbit outside of the IPD, a factor of 2-13 increase in sensitivity and a factor of 4-160 increase in observing speed can be achieved with no increase in telescope aperture. The exact improvement factor depends on the wavelength of the observation within the near-ultraviolet to far-infrared (UVOIR) spectrum. This advance will enable progress in space astrophysics using a mixture of large (flagship-class) and smaller, more affordable community proposed programs such as Explorer-class missions endowed with game-changing science capability afforded by available electric propulsion technology.

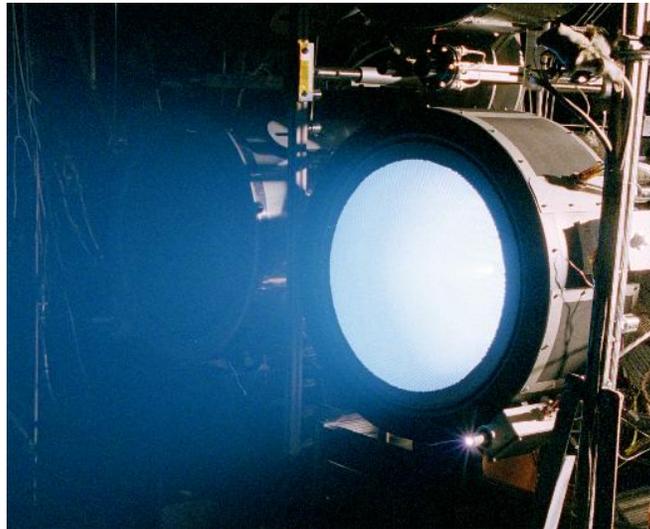

|  | NEXT | NSTAR | XIPS-25 | BPT-4000 |
|---|---|---|---|---|
| Thrust (mN) | 236 | 91 | 166 | 254 |
| Specific Impulse (s) | 4190 | 3070 | 3550 | 2150 |
| Input Power (kW) | 6.9 | 2.3 | 4.2 | 4.5 |
| Xenon Throughput (kg) | 300 - 500 | 155 - 200 | 150 | 275 - 300 |

Figure 1. NASA's Evolutionary Xenon Thruster (NEXT) uses electrical energy produced by solar panels to accelerate Xenon atoms producing thrust in a way that is much more efficient than conventional chemical thrusters. NEXT has demonstrated lifetime in ground testing that is 2X that required for the EZE mission architecture (top). Performance properties achieved by NEXT along with those of other SEP thrusters (bottom).

Astronomers search the world over for the darkest mountain tops on which to construct major ground-based observatories. Choosing a good telescope site is the most cost effective way to maximize the science capability of an observatory. So it is with space observatories. Until recently, limitations of in-space propulsion technology have prevented astronomers from selecting deep-space observing sites that enable the best observatory performance. NASA's investment in NEXT and other electric propulsion technologies can now enable space astronomers to realize "dark time" performance for the first time.

During 2011, we completed a detailed mission architecture study which demonstrated that an upper stage SEP propulsion module can be used to augment Falcon-9 launch services such that future astrophysics Explorer-class missions can escape the glow of the IPD[2]. We showed that Falcon-9, augmented by a 13 kW NEXT upper stage, can place a notional 700 kg Explorer payload into a 2

AU radius heliocentric orbit that is inclined by 15 degrees with respect to the ecliptic plane. This orbit is one of many possible choices and is sufficient to realize the above large gain in science performance relative to the LEO orbit of the Hubble Space Telescope, the Earth-trailing orbit of the Spitzer and Kepler Space Telescopes, or the Sun-Earth L2 point orbit of the James Webb and Herschel Space Telescopes. This study demonstrated that all enabling technologies for this propulsion module currently meet technology readiness requirements for infusion into a medium Explorer mission solicitation.

## 2. CREATING OPPORTUNITY FOR ALL EXPLORER PROGRAM PROPOSERS

Development of a standardized solar electric propulsion upper stage module for Falcon-9 (Figures 2 and 3), that all Astrophysics Explorer Program proposers can use, will yield a cost-efficient new capability that will set this small payload program on a new path to achieve major science goals, rivaling the capability of larger, more expensive systems. For this reason, we modeled our architecture study around a generic medium Explorer-class astrophysics payload rather than any specific mission proposed to the astrophysics decadal survey or elsewhere. Typical astrophysics science payload requirements such as telescope pointing control and high science data volume transmission were included, and an Explorer payload spacecraft bus was designed to enable a comprehensive study of the SEP module / science payload interface[2].

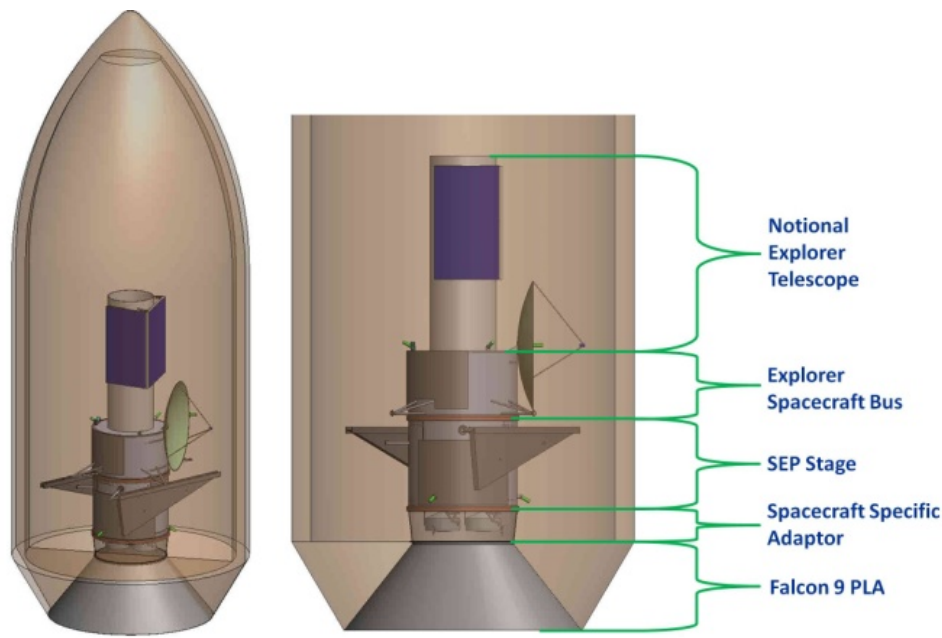

Figure 2. A notional 700 kg astrophysics Explorer payload mated with a 13 kW NEXT solar electric propulsion upper stage orbit transfer module is shown in a Falcon-9 fairing.

For compatibility with low cost payload programs such as Explorers and Discovery class missions, we developed a mission architecture approach in which the SEP capability was incorporated as an orbit transfer module upper stage that would be provided as an augmentation to launch services for community proposed science payloads. A clean interface approach was developed to enable program solicitations to treat the SEP module as residing on the launch vehicle side of interface to the proposed science payload. Command and control of the transfer module is provided by the science payload spacecraft to yield a module design with low recurring cost for repeated application to SMD astrophysics and planetary missions. To ensure compatibility with payloads requiring low pointing jitter, the transfer module is designed to eject upon orbit insertion leaving the payload free of large low frequency solar array structures. A conventional K-band RF system was designed to enable a daily uncompressed science data volume transmission of 40 – 380 Gbits to the DSN 36m ground station. The range in data volume depends upon position in the 2 AU 15 deg orbit.

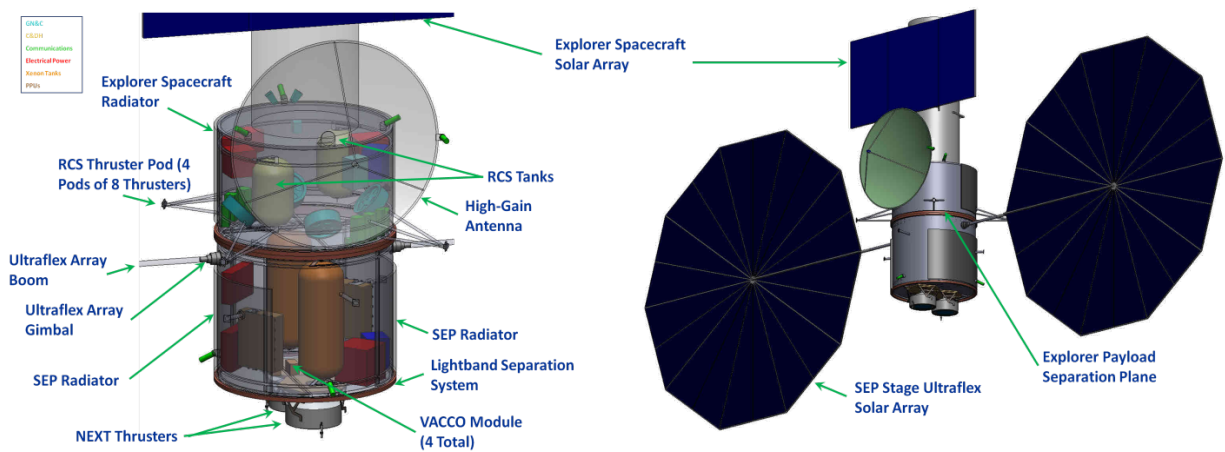

Figure 3. A clean interface between the SEP stage and the Explorer spacecraft bus has been developed to enable low technical and programmatic complexity (left); A notional Explorer-Class observatory with SEP upper stage shown in SEP cruise configuration with Ultraflex solar arrays deployed (right). The SEP stage and Ultraflex arrays eject upon science orbit insertion via a Lightband separation system leaving the science payload free of low modal frequency structure.

## 3. ACHIEVING MAJOR SCIENCE GOALS WITH SMALL APERTURE OBSERVATORIES

Cost constraints on space science necessitate maximizing the performance of small payload programs that can provide frequent access to space by university research teams. Making dark sky orbits accessible at low cost is a necessary step in ensuring the long-term scientific viability of the astrophysics Explorer program and will serve as a key first step toward enabling UVOIR flagship successors to the JWST to achieve optimal science performance per unit telescope aperture.

As shown in Figure 4, a substantial reduction in Zodiacal background power can be achieved by insertion into a modestly inclined 2 AU orbit such as that shown in Figure 5. The reduction in noise associated with this reduced background power enables broad-band imaging and slit-less spectroscopy systems working within the near-UV to far-infrared spectrum to realize a substantial increase in science performance shown in Figure 6. This advantage can be used to achieve higher science performance than would be possible using the same system in a more conventional orbit, or it can be used to achieve given science requirements with a smaller less expensive telescope aperture.

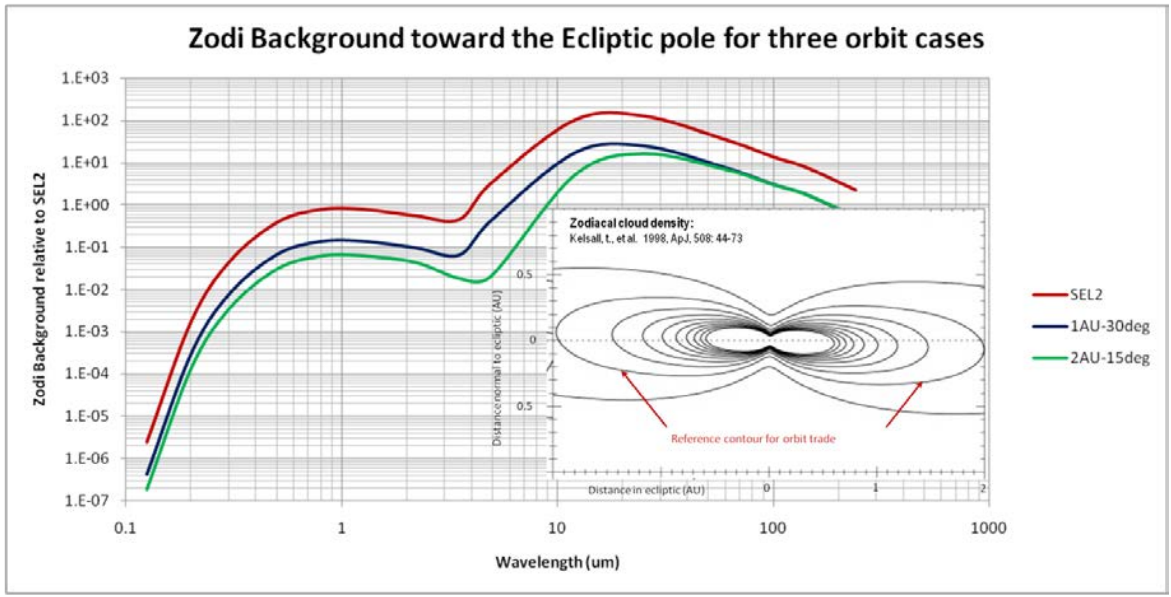

Figure 4: The Zodiacal background in two inclined heliocentric orbits derived from Kelsall et al.[1] is shown relative to a zero inclination orbit at the Sun-Earth L2 point. Inset: Isodensity contours of the IPD. Both orbits shown in the main figure extend outside of the reference contour shown.

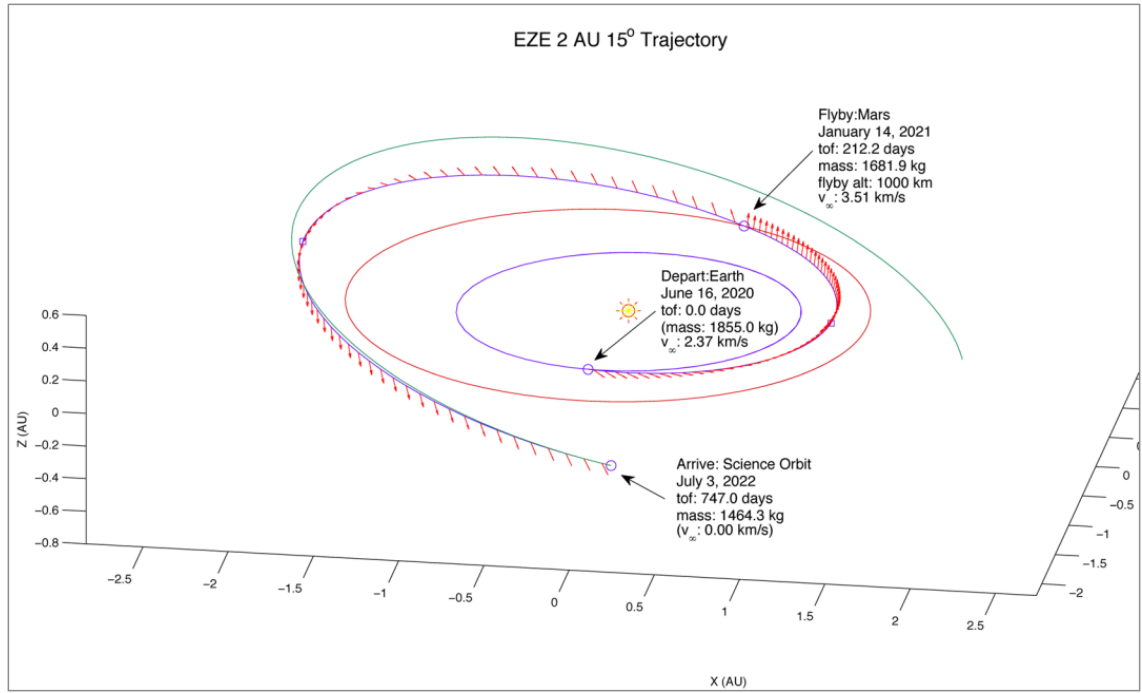

Figure 5: One of several candidate circular orbits modeled by the EZE mission architecture study[2] for a 700 kg Explorer science payload and Falcon-9 launch vehicle incorporating a 13 kW SEP orbit transfer module. Approximately half of the 15 deg orbit inclination is achieved by Mars gravity assist. The 2 year transfer time enables one full period of the 2 AU circular science orbit to occur within a 5 year duration mission. Thus any portion of the sky can be observed without looking back through the IPD.

Path finding dark sky operational science orbits with the above Explorer-scale approach offers immediate scientific return on investment and a substantial increase in observatory performance for all future UVOIR missions. The performance factors shown in Figure 6 result from reduced Zodiacal background power alone. However, when this reduction is combined with recent advances in photon counting detectors that achieve near noiseless detection of light[4], the resulting performance can be transformative for space astronomy – yielding performance per unit telescope aperture that can enable an increased role for small and medium aperture systems in development of space astrophysics strategy and policy.

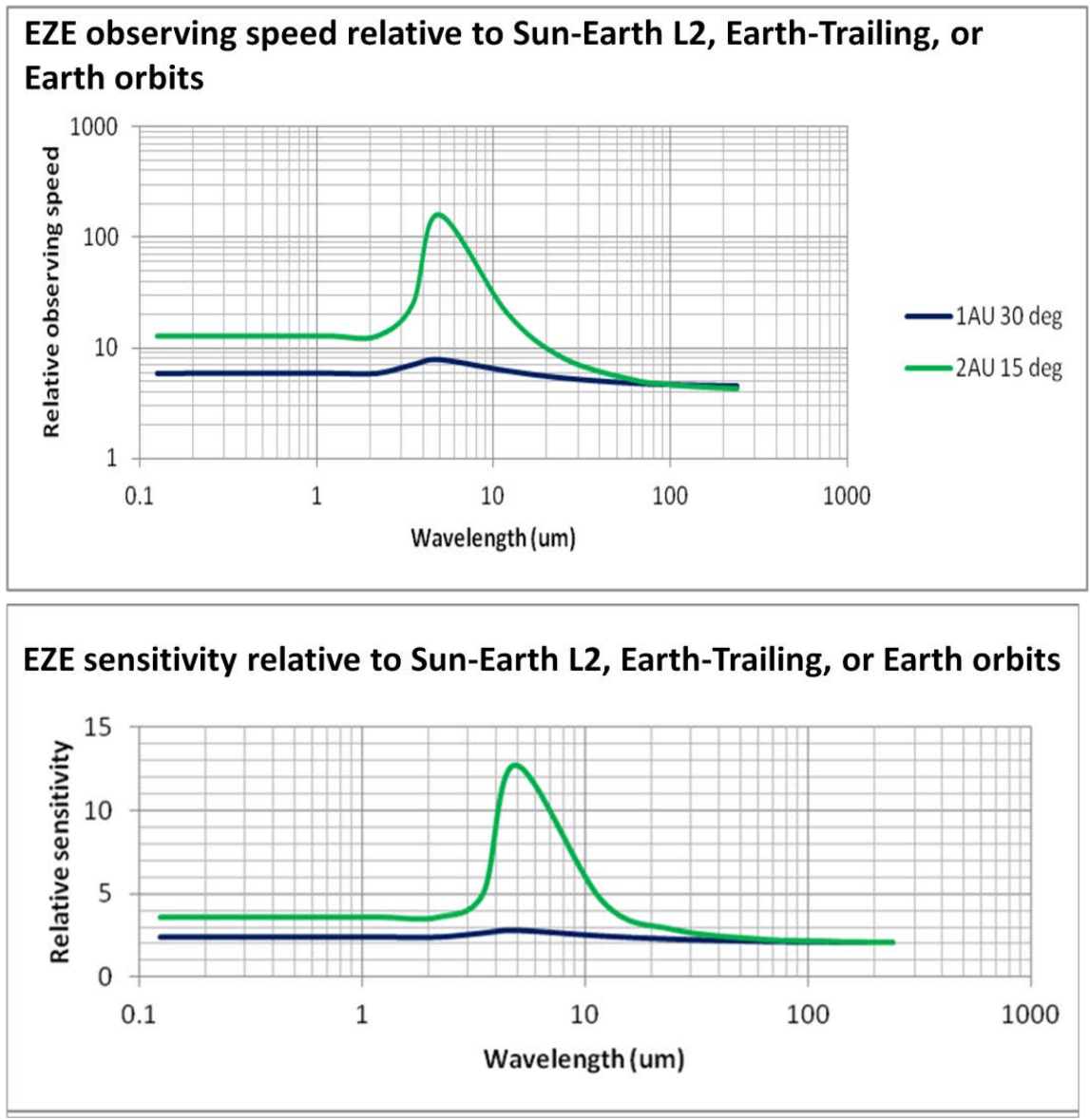

Figure 6: The relative survey speed and sensitivity performance gain for two heliocentric orbit cases: 1 AU 30 deg (blue) and 2 AU 15 deg (green).

# 4. PATHFINDING FOR COST EFFECTIVE UVOIR SPACE ASTRONOMY

NASA's need for further development of SEP capability cross-cuts both human and robotic exploration objectives as well as cross-cutting both astrophysics and planetary science disciplines within SMD. The Dawn mission[3] is the only NASA science mission to utilize SEP. As shown in Figure 7, the SEP module studied under EZE for astrophysics Explorers would realize a 7X increase in SEP thruster power over Dawn, and incorporates advances that are key to achieving flight systems that can scale to higher power for other NASA exploration roadmap objectives[5].

We note that infusing a SEP upper stage launch service augmentation into the next medium astrophysics Explorer mission solicitation would yield a program element that can potentially be synergized with existing OCT SEP flight demonstration plans to produce a flight mission that both advances this key pull technology while yielding immediate scientific return on investment. By adopting a SEP module design that is intended for duplication to support a wide range of SMD astrophysics and planetary missions, recurring costs would be minimized to yield an asset that can support a wide range of community proposed Explorer and Discovery class missions.

|  | EZE | Dawn |
| --- | --- | --- |
| Ion Thrusters | Two 7 kW NEXT for mission performance – parallel operation | Three 2.3 kW NSTAR – serial operation – one for fault tolerance |
| Power Processing Units | Two NEXT PPUs – parallel operation | Two NSTAR PPUs – one for fault tolerance |
| Xenon Tanks | Multiple off-the-shelf tanks | Single custom tank |
| Xenon Feed System | Proportional flow control-based system developed under NEXT technology | Bang-bang pressure regulation system derived from Deep Space 1 baseline |
| Digital Control Interface Unit | Functions incorporated into SEP stage Remote Interface Unit | Two custom-developed DCIUs – one for fault tolerance |
| Thruster Gimbals | NEXT technology-based | Custom developed for Dawn/NSTAR |

Figure 7: SEP features of the EZE orbit transfer module are compared to those of the DWAN spacecraft.

Development of complete flight-ready SEP systems, as discussed in the Planetary Decadal Survey[6], and consolidation of separate technology development efforts that cross-cut robotic and human exploration objectives, as recommended by the recent National Research Council review of NASA technology development roadmaps[7], can set SMD science on a new cost effective path of discovery within this decade. Consolidation of SEP technology efforts to yield systems, such as the EZE transfer module, that are applicable to near-term robotic mission objectives can enable a portfolio of SMD science missions that are diverse in scope and resilient to downward budget pressure on space science. This consolidated approach can enable steady flight demonstrated progress toward very high power SEP systems that are needed to enable human exploration objectives while producing near-term and continuous scientific return on investment.